\documentclass[structabstract]{aa}

\usepackage{graphicx}
\usepackage{rotating}
\usepackage{txfonts}
\def\dsct{$\delta$~Scuti}

\def\Msun{$M_{\odot}$}
\def\Lsun{$L_{\odot}$}
\def\Rsun{$R_{\odot}$}

\def\Teff{\ensuremath{T_{\mathrm{eff}}}}
\def\cd{d$^{\rm -1}$}
\def\logg{\ensuremath{\log g}}
\def\vmic{$\upsilon_{\mathrm{mic}}$}

\def\vsini{\ensuremath{{\upsilon}\sin i}}
\def\kms{$\mathrm{km\,s}^{-1}$}

\def\llm{{\sc LLmodels}}

\def\synth{{\sc SYNTH3}}

\begin{document}

\title{Regular frequency patterns in the young \dsct\, star HD~261711 observed by the CoRoT \thanks{The CoRoT  space mission was developed and is operated by the French space agency CNES, with participation of ESA's RSSD and Science Programmes, Austria, Belgium, Brazil, Germany, and Spain.} and MOST\thanks{Based on data from the {\it MOST} satellite, a Canadian Space Agency mission, jointly operated by Microsatellite Systems Canada Inc. (MSCI), formerly part of Dynacon, Inc., the University of Toronto Institute for Aerospace Studies and the University of British Columbia with the assistance of the University of Vienna.} satellites}

\author{K. Zwintz\inst{1,2}\thanks{Pegasus Marie Curie post-doctoral fellow of the Research Foundation - Flanders} \and
L. Fossati\inst{3} \and
D. B. Guenther\inst{4} \and
T. Ryabchikova\inst{5} \and
A. Baglin\inst{6}  \and
N. Themessl\inst{2} \and
T. G. Barnes\inst{7} \and
J. M. Matthews\inst{8} \and
M. Auvergne\inst{6} \and
D. Bohlender\inst{9} \and
S. Chaintreuil\inst{6} \and
R. Kuschnig\inst{2} \and
A. F. J. Moffat\inst{10} \and
J. F. Rowe\inst{11} \and
S. M. Rucinski\inst{12} \and
D. Sasselov\inst{13} \and
W. W. Weiss\inst{2}
}

\offprints{K. Zwintz, \\ \email{konstanze.zwintz@ster.kuleuven.be}}

\institute{
    Instituut voor Sterrenkunde, K. U. Leuven, Celestijnenlaan 200D, 3001 Leuven, Belgium \\
    \email konstanze.zwintz@ster.kuleuven.be \and
   University of Vienna, Institute of Astronomy, T\"urkenschanzstrasse 17, 1180 Vienna, Austria  \and
   Argelander-Institut f\"ur Astronomie der Universit\"at Bonn, Auf dem H\"ugel 71, 53121 Bonn, Germany \and
   Department of Astronomy and Physics, St. MaryÕs University, Halifax, NS B3H 3C3, Canada \and
   Institute of Astronomy, Russian Academy of Sciences, Pyatnitskaya 48, 119017 Moscow, Russia \and
    LESIA, Observatoire de Paris-Meudon, 5 place Jules Janssen, 92195 Meudon, France \and
   The University of Texas at Austin, McDonald Observatory, 82 Mt. Locke Rd., McDonald Observatory, Texas 79734, USA \and
   Department of Physics and Astronomy, University of British Columbia,
    6224 Agricultural Road, Vancouver, BC V6T 1Z1, Canada \and
       Herzberg Institute of Astrophysics, National Research Council of Canada, 5071 West Saanich Road, Victoria, BC V9E 2E7, Canada \and
    D\'epartment de physique, Universit\'e de Montr\'eal, C.P.6128, Succ. Centre-Ville, Montr\'eal, QC H3C 3J7, Canada \and
    NASA-Ames Research Park, MS-244-30, Moffett Field, CA 94035, USA \and
    Department of Astronomy \& Astrophysics, University of Toronto, 50 St. George Street, Toronto, ON M5S 3H4, Canada \and
    Harvard-Smithsonian Center for Astrophysics, 60 Garden Street,
    Cambridge, MA 02138, USA
    }

\date{Received / Accepted }

\abstract
{The internal structure of pre-main-sequence (PMS) stars is poorly constrained at present. This could change significantly through high-quality asteroseismological observations of a sample of such stars. }
{We concentrate on an asteroseismological study of HD 261711, a rather hot \dsct-type pulsating member of the young open cluster NGC 2264 located at the blue border of the instability region.  HD 261711 was discovered to be a pre-main sequence \dsct\, star using the time series photometry obtained by the MOST satellite in 2006. }
{High-precision, time-series photometry of HD 261711 was obtained by the MOST and CoRoT satellites in four separate new observing runs that are put into context with the star's fundamental atmospheric parameters obtained from spectroscopy. Frequency Analysis was performed using Period04. The spectral analysis was performed using equivalent widths and spectral synthesis.}
{With the new MOST data set from 2011/12 and the two CoRoT light curves from 2008 and 2011/12, the \dsct\, variability was confirmed and regular groups of frequencies were discovered. The two pulsation frequencies identified in the data from the first MOST observing run in 2006 are confirmed and 23 new \dsct-type frequencies were discovered using the CoRoT data. Weighted average frequencies for each group were determined and are related to $l$ = 0 and $l$ = 1 $p$-modes. Evidence for amplitude modulation of the frequencies in two groups is seen. The effective temperature (\Teff) was derived to be 8600 $\pm$ 200\,K, \logg\, is 4.1 $\pm$ 0.2, and the projected rotational velocity (\vsini) is 53 $\pm$ 1\kms. Using our \Teff\, value and the radius of 1.8 $\pm$ 0.5 \Rsun\, derived from spectral energy distribution (SED) fitting, we get a luminosity log\,$L$/\Lsun\, of 1.20 $\pm$ 0.14 which agrees well to the seismologically determined values of 1.65 \Rsun\, and, hence, a log\,$L$/\Lsun\, of 1.13. The radial velocity of 14 $\pm$ 2\,\kms\,we derived for HD 261711, confirms the star's membership to NGC 2264.}
{Our asteroseismic models suggest that HD 261711 is a \dsct-type star close to the zero-age main sequence (ZAMS) with a mass of 1.8 to 1.9\Msun. With an age of about 10 million years derived from asteroseismology, the star is either a young ZAMS star or a late PMS star just before the onset of hydrogen-core burning. The observed splittings about the $l$ = 0 and 1 parent modes may be an artifact of the Fourier derived spectrum of frequencies with varying amplitudes. }


\titlerunning{The (PMS) $\delta$ Scuti star HD 261711}
\authorrunning{K. Zwintz et al.}
\maketitle

\section{Introduction}

Asteroseismology is an important science goal in the two space missions, MOST (Walker et al. \cite{wal03}) and CoRoT (Baglin \cite{bag06}). Numerous publications on different asteroseismic targets reflect the importance of observations conducted with these two space telescopes. Pulsating pre-main sequence (PMS) stars have been successfully studied with both satellites as well (e.g. Zwintz et al. \cite{zwi11a}). 
The young open cluster NGC 2264, which is rich in young stellar objects, is located in a region accessible to both telescopes and, indeed, the cluster has been observed twice by MOST and twice by CoRoT.
Among the common targets of the two satellites in NGC 2264 is the young \dsct\, star HD 261711, the subject of this study.

NGC 2264 ($\alpha_{2000}$ = $6^h$ $41^m$, $\delta_{2000}$ = $+9^{\circ}$ $53'$) was studied frequently in the past using various instruments in different wavelength ranges from space and from the ground.
The cluster is located in the Monoceros OB1 association about 30\,pc above the galactic plane and has a diameter of $\sim$39 arcminutes. Sung et al. (\cite{sun97}) report a cluster distance of 759 $\pm$ 83pc which corresponds to a distance modulus of 9.40 $\pm$ 0.25 mag. 
Kharchenko et al. (\cite{kha01}) find proper motion values of -2.70 $\pm$ 0.25 mas/yr in right ascension and -3.50 $\pm$ 0.26 mas/yr in declination for NGC~2264. The mean value of the cluster reddening is rather low and lies at $E(B-V)=0.071 \pm 0.033$ mag (Sung et al. \cite{sun97}); the differential reddening across NGC 2264 is negligible.  

The age of NGC\,2264 can only be determined with a relatively large error as its main sequence consists only of massive O and B stars and stars of later spectral types that are still in their pre-main sequence phase. Therefore, different values for the cluster's age are reported in the literature; they range from 3  to 10 million years (e.g., Sung et al. \cite{sun04}, Sagar et al. \cite{sag86}).

The $V=11.3$ mag bright star HD~261711 (NGC~2264~39; GSC~00746-01783) has a spectral type of A2V (Skiff \cite{ski05}) and a $(B-V)$ value of 0.13 mag (including a reddening of $E(B-V)=0.071$ mag as given by Sung et al. \cite{sun97}). According to these parameters, HD~261711 falls into the (PMS) instability strip for \dsct-type pulsation.
The star is located in the outer part of NGC 2264 at $\alpha_{2000}$ = $6^h$ $40^m$ $16.18^s$, $\delta_{2000}$ = $+9^{\circ}$ $17'$ $13.14'$, but with proper motion values of -3.00 $\pm$ 2.60 mas/yr in right ascension and -4.20 $\pm$ 2.40 mas/yr in declination (H{\o}g et al. \cite{hog00}) it is likely to be a cluster member. 

During their evolution from the birthline to the zero-age main sequence (ZAMS), PMS stars cross different instability regions in the Hertzsprung-Russell (HR)-diagram, hence can become vibrationally unstable (e.g., Marconi \& Palla \cite{mar98}, Bouabid et al. \cite{bou11}). Intermediate mass PMS stars with spectral types from A to early F are frequently found to show \dsct-like pulsation (e.g., Zwintz \cite{zwi08}). 

In the past, extensive ground- and space-based photometric campaigns have discovered many hundreds of ``classical'' (post-) main sequence \dsct\, stars. \dsct-type pulsation frequencies are typically in the range between about 5\cd\, and up to 80\cd, and have $p$-modes driven by the $\kappa$-mechanism operating in the H, He I and He II ionization zones. Statistically, the photometrically observed modes are not distributed randomly, but cluster around the frequencies of the radial modes over many radial orders (Breger et al. \cite{bre09}). Such regular frequency spacings might be explained, for example, by modes trapped in the stellar envelope (Breger et al. \cite{bre08}) or by combination modes (Breger et al. \cite{bre11}).

For asteroseismic modeling the identification of the modes is essential, but very often this is only possible for few of the observed frequencies. Therefore, the recognition of regularities in the observed pulsation frequency spectra is an important additional tool. Recently, several pre- and (post-) main sequence \dsct\, stars which show regular frequency patterns have been discovered, such as HD 144277 (Zwintz et al. \cite{zwi11b}), KIC9700322 (Breger et al. \cite{bre11}) and HD 34282 (Casey et al. \cite{cas13}).

In 2006, the MOST space telescope observed NGC~2264 for the first time with the purpose to search for pulsating PMS stars.
Indeed, four new PMS \dsct\ stars were revealed and the two known PMS pulsators, V~588~Mon and V~589~Mon, were confirmed (Zwintz et al. \cite{zwi09}). 
HD~261711 is among the four PMS pulsators discovered then and was denoted as ``V2" in Zwintz et al. (\cite{zwi09}). Two \dsct-type pulsation frequencies were revealed for HD~261711 in the 2006 MOST data (Zwintz et al. \cite{zwi09}).

In this study we combine the already published time series photometry of HD~261711 obtained with the MOST satellite in 2006 with recent MOST observations conducted in 2011/12 and with high-precision data from the CoRoT satellite observed in 2008 and 2011/12. We use a high-resolution spectrum to determine the atmospheric parameters and chemical abundances of HD~261711,  investigate the star's location in the HR-diagram, conduct an asteroseismic investigation of the observed frequencies and discuss HD 261711's evolutionary stage.

\begin{table}[htb]
\caption{Characteristics of the MOST and CoRoT observations of HD~261711. }
\label{obs}
\begin{center}
\begin{tabular}{lrrcrc}
\hline
\hline
\multicolumn{1}{l}{data set} & \multicolumn{1}{c}{data points} & \multicolumn{1}{c}{tbase} & \multicolumn{1}{c}{1/T} & \multicolumn{1}{c}{f$_{\rm Nyquist}$} & \multicolumn{1}{c}{noise$_{\rm res}$} \\
\multicolumn{1}{c}{ } & \multicolumn{1}{c}{\#}  & \multicolumn{1}{c}{[d]} & \multicolumn{1}{c}{[d$^{-1}$]} & \multicolumn{1}{c}{[d$^{-1}$]} & \multicolumn{1}{c}{[mmag]}\\
\hline
MOST06 & 10531 & 22.72 & 0.044 &  1362.23 & 0.752\\
MOST11/12 & 8391 & 39.98 & 0.025 & 705.40 & 0.757 \\
CoRoT08 & 49345 & 20.81 & 0.048 & 1337.21 & 0.014 \\
CoRoT11/12 & 91168 & 38.69 & 0.025 & 1221.31 & 0.009 \\
\hline
\end{tabular}
\end{center}
\tablefoot{Data set, number of data points used for the analysis, time base, Rayleigh frequency resolution 1/T, Nyquist frequency and noise level of the residuals (noise$_{\rm res}$).}
\end{table}

\section{MOST observations and data reduction}

The MOST space telescope (Walker et al. \cite{wal03}) was launched on 30 June 2003 into a polar Sun-synchronous circular orbit. At an altitude of 820\,km it revolves around the Earth with a period of 101.413\,minutes. The satellite carries a 15-cm Rumak-Maksutov telescope which feeds a CCD photometer through a single, custom broadband filter (wavelength range from 3500 to 7500\,\AA). MOST has successfully observed numerous objects over the past 9 years of its continuing operation.

\begin{figure}[htb]
\centering
\includegraphics[width=0.5\textwidth,clip]{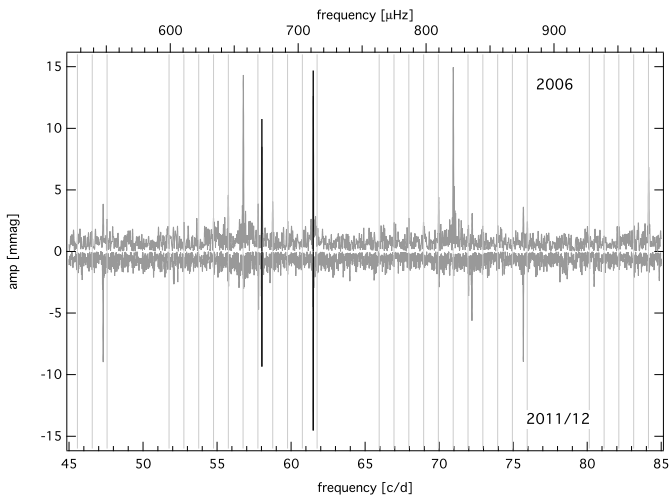}
\caption{MOST data for HD 261711: amplitude spectra from 45 to 85 c/d obtained in 2006 (top panel) and 2011/12 (bottom panel). The top X axes give the frequencies in $\mu$Hz.}
\label{most0611}
\end{figure}

Three types of photometric data are provided simultaneously by the MOST space telescope for different targets in its field of view. Data obtained in Fabry Imaging Mode are generated by a projection of the entrance pupil of the telescope -- illuminated by a bright ($V < 6$ mag) target star -- onto the Science CCD by a Fabry microlens (see Reegen et al. \cite{ree06} for details). Direct Imaging is used for stars in the open area of the CCD which is not covered by the Fabry microlens array field stop mask. It resembles conventional CCD photometry where photometry is obtained from defocussed images of stars. Although originally not intended for scientific purposes, the MOST Guide Stars used for the Attitude Control System (ACS) provide highly accurate photometry which has been frequently used in the past (e.g., Zwintz et al. \cite{zwi09}). 

MOST observed NGC~2264 including its member HD~261711 for the first time from December 7, 2006, to January 3, 2007, in a dedicated observing run on the cluster itself (Zwintz et al. \cite{zwi09}). 
The second MOST observing run on NGC~2264 lasted from December 5, 2011, to January 14, 2012, and was conducted as part of the CSI NGC 2264 (Coordinated Synoptic Investigation of NGC 2264) project together with the space telescopes CoRoT (Baglin \cite{bag06}), Spitzer (Werner et al. \cite{wer04}) and Chandra (Weisskopf et al. \cite{wei02}). Note that HD 261711 was not observed by the Spitzer and Chandra satellites during this campaign, i.e., only measurements in the optical from MOST and CoRoT are available.

Due to the magnitude range (7 $< V <$ 12 mag) and the large number of targets, in both runs NGC~2264 was observed in the open field of the MOST Science CCD in Guide Star Photometry Mode. Because not all of the candidate PMS pulsators (i.e., A and F type cluster stars) could be reached using a single pointing of the satellite, two fields of observations were chosen and observed in alternating halves of each 101-min orbit. 
Using this setting, MOST time series photometry was obtained for a total of 68 stars in the region of NGC 2264 in the 2006 run (Zwintz et al. \cite{zwi09}) and 67 in the 2011/12 run.

The corresponding MOST 2006 light curve of HD~261711 used for the analysis has a time base of $\sim$22.7\,d. The on-board exposure time was 1.5\,s, 16 consecutive images were ``stacked" on board resulting in an integration time of 24\,s (see also Zwintz et al. \cite{zwi09}).
The total length of the HD~261711 data set from the 2011/12 run is 39.98\,d. In 2011/12  on-board exposures were 3.01\,s long (to satisfy the cadence of guide star ACS operations), and the integration time was 51.17\,s as 17 consecutive images were ``stacked" on board.

Data reduction of the MOST Guide Star photometry was conducted using the method developed by Hareter et al. (\cite{har08}) which uses a similar approach as for targets observed in Fabry imaging mode (Reegen et al. \cite{ree06}), i.e., resolving linear correlations between the intensity of the target and background pixels. Since MOST Guide Star photometry does not provide information on the intensity of the background, the correlations between constant and variable stars are used instead. In a first step, the MOST Guide Stars are classified into variable and constant objects by a quick-look analysis. The data of all selected intrinsically constant objects are then combined to a comparison light
curve. 

Stray light effects are corrected by subtraction of linear correlations between target and comparison time-series. It is selfevident that comparison time-series are assumed and required to contain no variable stellar signal. If one of the stars used for the comparison light curve turns out to be variable (even at low amplitude levels) in a later stage of the analysis, the complete reduction has to be repeated omitting the variable star. After the Guide Star photometry reduction, the 2006 light curve of HD~261711 consists of 10531 data points, and the 2011/12 light curve has 8391 data points corresponding to Nyquist frequencies of 1362.23\,\cd\ and 705.40\,\cd, respectively. Note that the 2011/12 light curve has fewer data points than the 2006 light curve although the time base is significantly longer. The reason is that in 2011/12 HD~261711 was observed in the part of the orbit affected by worse stray light conditions than in 2006, hence more data points had to be discarded in the reduction.

An overview of the properties of the MOST observations is given in Table \ref{obs}. The respective amplitude spectra from 2006 and 2011/12 are shown in Figure \ref{most0611}. Note that the peaks between 70 and 75 \cd\ and between 45 and 50\cd\ are alias frequencies to F1 and F2 with the MOST orbital frequency and disappear after prewhitening F1 and F2.

\begin{figure}[htb]
\centering
\includegraphics[width=0.48\textwidth,clip]{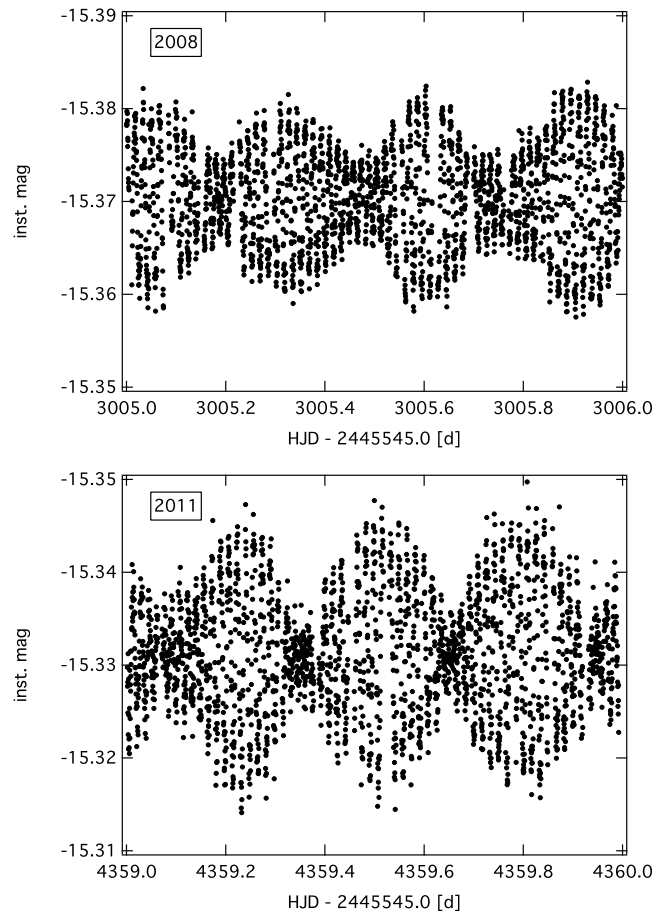}
\caption{One-day subsets of the CoRoT light curves obtained in 2008 (top panel) and 2011/12 (bottom panel) to the same scales.}
\label{lcs}
\end{figure}

\begin{figure*}[htb]
\centering
\includegraphics[width=0.95\textwidth,clip]{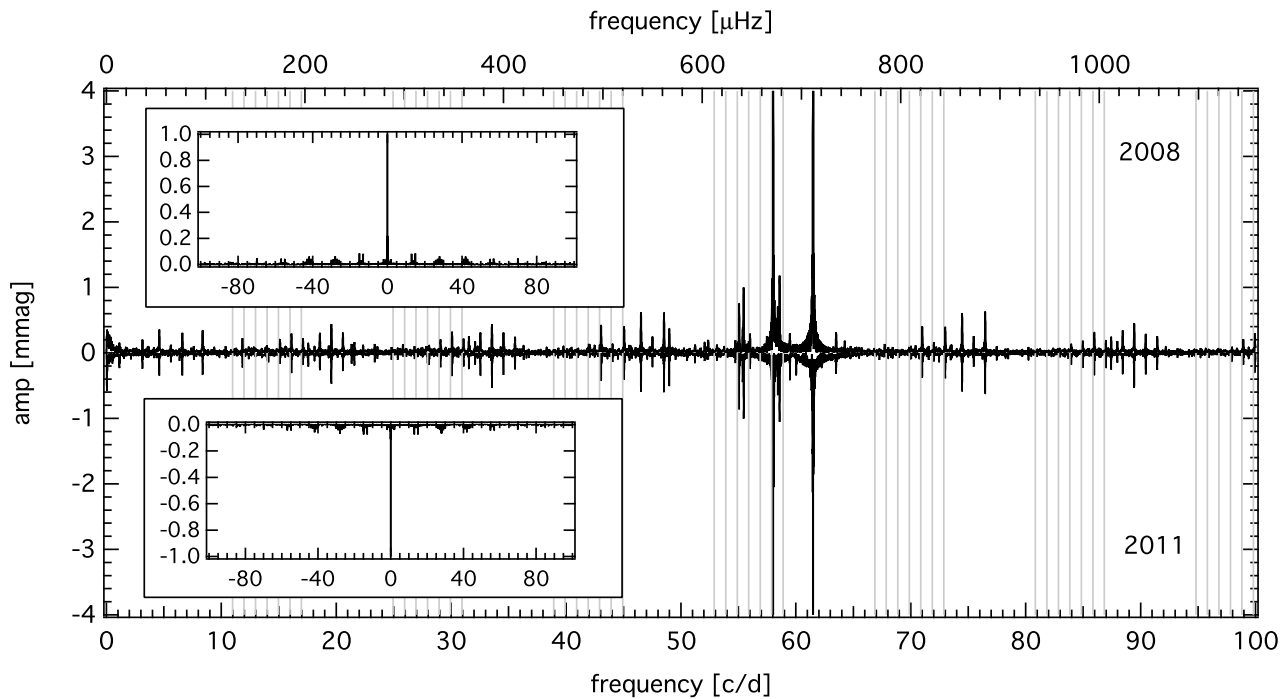}
\caption{Original amplitude spectra of the CoRoT data obtained in 2008 (pointing upwards) and 2011/12 (pointing downwards) from 0 to 100\,\cd\ (i.e., from 0 to 1157.4\,$\mu$Hz) and zoomed in amplitude to 4\,mmag. The corresponding spectral window functions are given as inserts, respectively.}
\label{corotorigamps}
\end{figure*}

\section{CoRoT observations and data reduction}

The CoRoT satellite (Baglin \cite{bag06}) was launched on December 27, 2006, from Baikonur aboard a Soyuz rocket into a polar, inertial circular orbit at an altitude of 896\,km. 
CoRoT carries a 27-cm telescope and can observe stars inside two cones of 10$^{\circ}$ radius, one at $RA=06:50$ and the other at $RA=18:50$. The field of view of the telescope is almost circular with a diameter of 3.8\,$^{\circ}$ and the filter bandwidth ranges from 3700 to 10000\,\AA.

CoRoT observed NGC\,2264 for the first time for 23.4\,d in March 2008 during the Short Run SRa01 within the framework of the {\it Additional Programme} (Weiss \cite{wei06}). A second run on the cluster NGC\,2264 (SRa05) was conducted in December 2011 / January 2012 with a time base of about 39\,d as part of the previously mentioned 
CSI NGC 2264 project including also the satellites MOST, Spitzer and Chandra.

For both observing runs, the complete cluster was placed in one Exofield CCD and data were taken for all stars in the accessible magnitude range, i.e., from 10 to 16\,mag in $R$. The 100 brightest stars in the field of NGC\,2264 were  primary targets to search for stellar pulsations among PMS cluster members.

The reduced N2 data for HD~261711 were extracted from the CoRoT data archive. The CoRoT data reduction pipeline (Auvergne et al. \cite{auv09}) flags those data points that were obtained during passages of the satellite over the South Atlantic Anomaly (SAA). We did not use these `SAA-flagged' data points in our analysis. 
The light curve of HD~261711 obtained in 2008 consists of 49345 data points with a sampling time of 32\,s. The 2011/12 data set of HD~261711 has been again observed with a sampling time of 32\,s and comprises 91168 data points. One-day subsets of the CoRoT light curves from 2008 and 2011/12 are shown in Figure \ref{lcs}.
Table \ref{obs} summarizes the characteristics of the CoRoT light curves.

\begin{figure*}[htb]
\centering
\includegraphics[width=0.95\textwidth,clip]{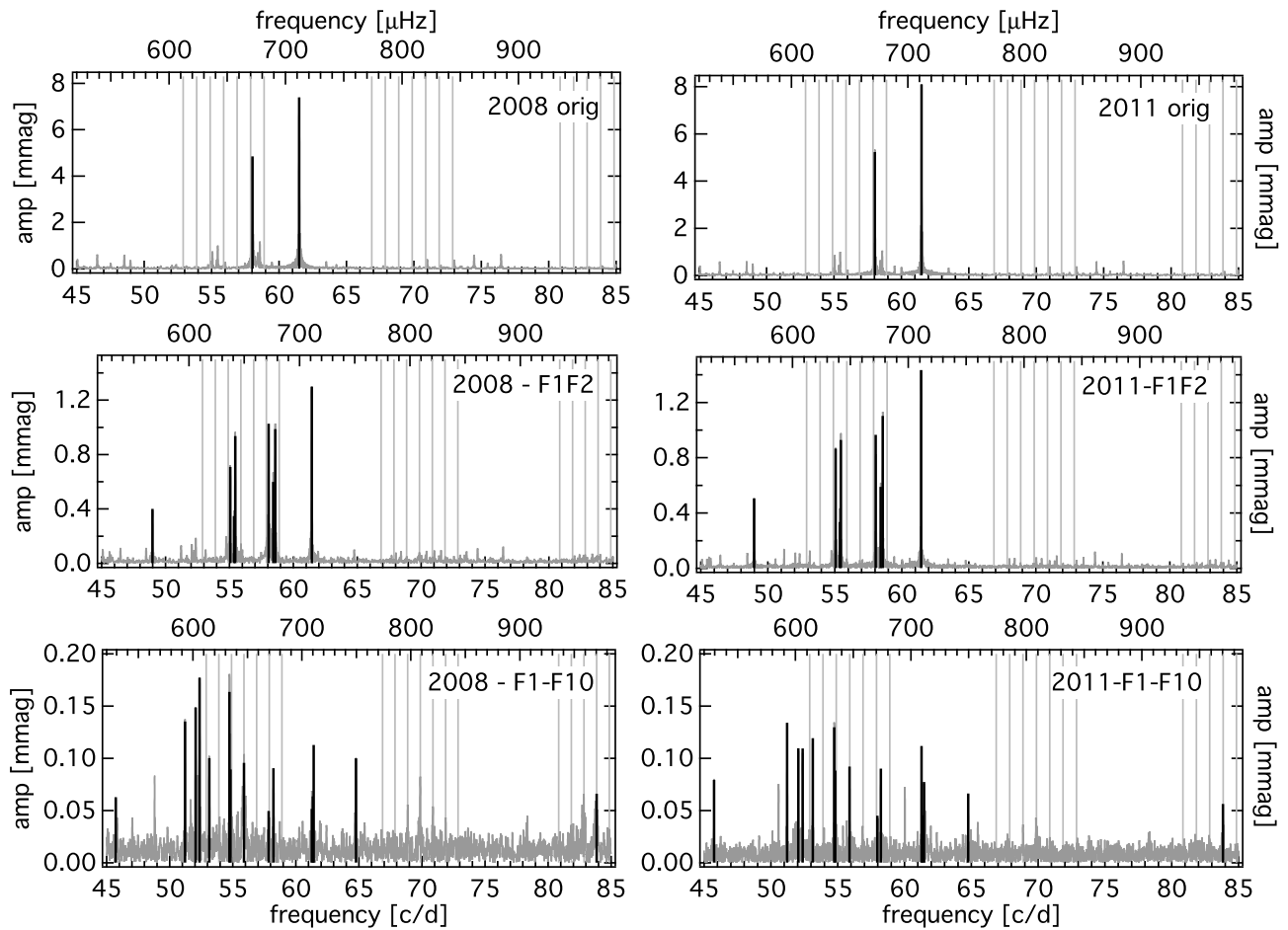}
\caption{CoRoT data for HD 261711: original amplitude spectra from 45 to 85 c/d obtained in 2008 (left) and 2011/12 (right) with F1 and F2 marked, residual amplitude spectra after prewhitening F1 and F2 where F3 to F10 are marked (middle panels) and residual amplitude spectra after prewhitening F1 to F10 showing F11 to F25. The top X axes give the frequencies in $\mu$Hz.}
\label{corotamps}
\end{figure*}

\section{Frequency Analysis}\label{freqana}

For the frequency analyses of all four data sets, we used the software package Period04 (Lenz \& Breger \cite{len05}) that combines Fourier and least-squares algorithms. Frequencies were then prewhitened and considered to be significant if their amplitudes exceeded four times the local noise level in the amplitude spectrum (i.e., 4 S/N; Breger et al. \cite{bre93}, Kuschnig et al. \cite{kus97}). 

We verified the analysis using the SigSpec software (Reegen \cite{ree07}). SigSpec computes significance levels for amplitude spectra of time series with arbitrary time sampling. The probability density function of a given amplitude level is solved analytically and the solution includes dependences on the frequency and phase of the signal.

Each data set was analyzed independently and the results were then compared to each other. 

The two MOST light curves of HD~261711 were obtained only during half-orbits and HD~261711 with a $V$ magnitude of 11.3 is a rather faint target for the MOST space telescope. 
In the frequency analysis of the MOST light curve from 2006, only two frequencies were attributed to pulsation; the other peaks are connected to the orbital frequency of the satellite, its harmonics and 1~\cd\ sidelobes (Zwintz et al. \cite{zwi09}). These two pulsation frequencies, i.e., F1 at 61.498\,$\pm$0.002~\cd\  (711.78\,$\pm$0.02\,$\mu$Hz) and F2 at 58.027\,$\pm$0.003~\cd\  (671.60\,$\pm$0.04\,$\mu$Hz) are confirmed in the 2011/12 data set (see Figure \ref{most0611} and Table \ref{freqs}) and no additional peaks originating from pulsation could be found.
The residual noise levels for the two MOST light curves from 2006 and 2011/12 after prewhitening all significant frequencies are 0.752 and 0.757\,mmag, respectively.

\begin{figure}[htb]
\centering
\includegraphics[width=0.48\textwidth,clip]{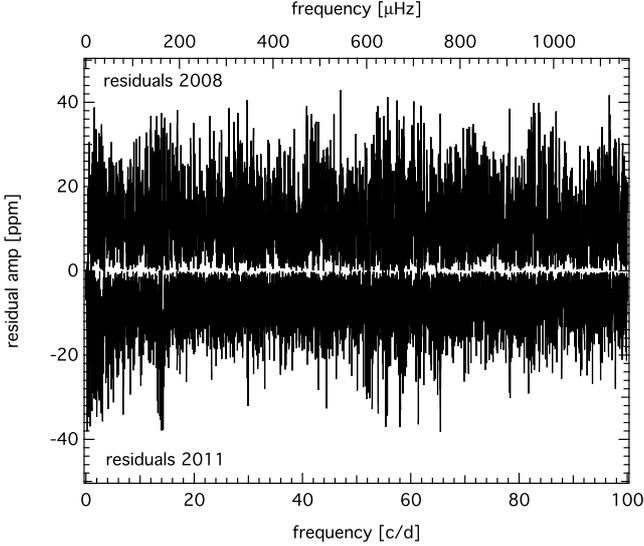}
\caption{Residuals of the CoRoT data from 2008 (pointing upwards) and 2011/12 (pointing downwards) after prewhitening all significant frequencies. Note that the Y axis is given in ppm.}
\label{corotresiduals}
\end{figure}

In the CoRoT photometry, the influence of the satellite's orbital and related frequencies has to be taken into account as well, but the effects are less than for the MOST time series. The complete amplitude spectra for both years from 0 to 100\,\cd\ with a zoom in amplitude up to 4 millimagnitudes are given in Figure \ref{corotorigamps} with the corresponding spectral window functions given as insets. 
The frequency analyses of the CoRoT light curves from 2008 and 2011/12 yielded 25 common frequencies which were attributed to pulsation. The two frequencies found in the MOST data sets, F1 and F2, were confirmed using the CoRoT data. Table \ref{freqs} lists all frequencies together with the amplitudes derived from the different years and including the last digit errors given in parentheses computed according to Kallinger et al. (\cite{kal08}). Figure \ref{corotamps} shows zooms into the two amplitude spectra of the original data sets (top panels), residuals after prewhitening with the two highest amplitude frequencies, F1 and F2 (middle panels) and residuals after prewhitening the first ten frequencies. The corresponding 25 pulsation frequencies are identified as solid black lines. After prewhitening all significant frequencies the residual noise levels of the 2008 and 2011/12 data sets are 13.7\,ppm and 9.15\,ppm, respectively. The residuals are shown in Figure \ref{corotresiduals}.

Figure \ref{corotamps} illustrates the presence of regular patterns of groups of frequencies with nearly equidistant spacing. 
Similar effects have been observed in other \dsct-type stars, such as in HD 144277 (Zwintz et al. \cite{zwi11b}) and HD 34282 (Casey et al. \cite{cas13}). The pulsation frequencies of all three objects are rather high and with similar separations of the groups of observed frequencies lying between 3 and 4 \cd. All three stars are also hot objects close to the blue border of the instability region for (PMS) \dsct-type stars.

\begin{table*}[htb]
\caption{Results from the frequency analysis of HD 261711 using MOST and CoRoT data.} 
\label{freqs}
\begin{center}
\begin{scriptsize}
\begin{tabular}{llrrcrrlrrrr}
\hline
\hline
\multicolumn{1}{l}{data set} &\multicolumn{1}{l}{F} & \multicolumn{2}{c}{frequency} 
& \multicolumn{1}{c}{amp$_1$} & \multicolumn{1}{c}{S/N$_1$} & \multicolumn{1}{c}{sig$_1$}  & \multicolumn{1}{c}{amp$_2$} & \multicolumn{1}{c}{S/N$_2$} & \multicolumn{1}{c}{sig$_2$} 
& \multicolumn{1}{c}{amp$_2$ - amp$_1$ }& \multicolumn{1}{c}{lin. combi.} \\
\multicolumn{1}{l}{ } &\multicolumn{1}{l}{\#} & \multicolumn{1}{c}{[d$^{-1}$]}  & \multicolumn{1}{c}{[$\mu$Hz]}  & \multicolumn{1}{c}{[mmag]}  & \multicolumn{1}{c}{ }  & \multicolumn{1}{c}{ }  & \multicolumn{1}{c}{[mmag]}  & \multicolumn{1}{c}{ }  & \multicolumn{1}{c}{ } & \multicolumn{1}{c}{[mmag]}  & \multicolumn{1}{c}{ }\\
\hline
MOST & F1 & 61.498(2) & 711.78(2) & 14.817 & 18.6 & 127.9 & 14.525 & 20.3 & 144.9 & -0.292 & \\
               & F2 & 58.027(3) & 671.60(4) &  10.400 & 9.8 & 68.6 &	 9.335 & 11.5 & 62.9 & -1.065 & \\
\hline
CoRoT & F1 & 61.4976(2)  &	711.778(3)  &	7.392 & 535.9  & 6618.6    &	8.103 & 24.0 & 11786.7  &  0.711 &  \\
               & F2 & 58.0272(2)  & 671.611(3) &	4.850 & 319.6  & 7391.6    & 	5.230  &  63.8 & 12512.9   & 0.380 & \\
               & F3 & 61.4401(5) & 711.112(6) &   1.298 & 94.1 & 1661.9  & 1.435 & 83.0 & 2457.7  & 0.137 & \\
               & F4 & 58.5821(6) & 678.034(7) &  0.988 & 67.1 & 1311.8  & 1.103 & 19.3 & 1774.1  &  0.115 & \\
               & F5 & 58.0553(8) & 671.937(9) &  	1.027 & 66.7 & 1459.3  & 0.966 & 30.8 & 1126.7  & -0.061 & \\
               & F6 & 	55.4566(7) & 641.858(8) &  0.937 & 56.5 & 1330.7  & 0.928 & 19.1 & 1441.3 & -0.009 & \\
               & F7 & 	55.0608(7) & 637.278(9) & 0.710 & 42.9 & 928.6  & 0.866 & 33.1 & 1191.9  & 0.156 &  \\
               & F8 & 	58.429(1) & 676.26(1) &  0.599 & 39.7 & 740.1  & 0.589 & 42.9 & 644.4 & -0.010 &  \\
               & F9 & 	48.982(1) & 566.92(1) &  0.401 & 32.4 & 356.8  & 0.506 & 60.3 & 505.9 & 0.105 & \\
               & F10 & 	55.366(2) & 640.81(2) &  0.348 & 21.1 & 286.2  & 0.334 & 20.8 & 220.6  & -0.014 & F3+F9-F7  \\
               & F11 & 	51.217(4) & 592.79(5) &  0.135 & 10.2 & 50.8  & 0.134 & 9.7 & 35.8 &  -0.001 & \\
               & F12 & 	54.742(4) & 633.58(5) &  0.164 & 10.3 & 82.3  & 0.130 & 11.1 & 37.9  & -0.034 & F11+F4-F7 \\
               & F13 & 	53.131(5) & 614.94(6) &  0.100 & 6.7 & 29.1  & 0.119 & 9.1 & 28.5 & 0.019 & \\
               & F14 & 	61.281(5) & 709.27(6) &  0.052 & 3.7 & 16.3  & 0.112 & 8.5 & 25.3 & 0.060 & \\
               & F15 & 	52.057(5) & 602.51(6) &  0.149 & 10.4 & 65.3  & 0.110 & 8.3 & 24.4 & -0.039 & \\
               & F16 & 	52.374(5) & 606.18(6) &  0.177 & 12.9 & 83.9  & 0.109 & 10.2 & 25.3 & -0.068 & \\
               & F17 & 	55.891(6) & 646.89(7) &  0.096 & 5.9 & 24.8  & 0.092 & 8.8 & 16.5  & -0.004 & \\
               & F18 & 	58.231(6) & 673.97(7) &  0.091 & 5.9 & 29.9  & 0.090 & 7.4 & 16.6  & -0.001 & \\
               & F19 & 	54.808(6) & 634.35(7) &  0.089 & 5.6 & 19.0  & 0.088 & 8.5 & 15.9 & -0.001 & \\
               & F20 & 	45.739(7) & 529.39(8) &  0.063 & 4.8 & 10.9  & 0.079 & 8.8 & 13.5 & 0.016 & \\
               & F21 & 	61.463(7) & 711.38(9)&  0.113 & 8.2 &  8.4  & 0.077 & 6.8 & 11.8  & -0.036 & F3+F5-F2\\
               & F22 & 	64.764(8) & 749.6(1) &  0.100 & 7.9 & 27.3  & 0.066 & 6.5 & 9.5 & -0.034 & \\
               & F23 & 	83.84(1) & 970.3(1) & 0.066 &  4.4 & 8.6  & 0.056 & 6.6 & 6.6 & -0.010 & \\
               & F24 & 	61.34(1) & 709.9(1) &  0.063 &   4.6 & 8.1  & 0.046 & 4.4 & 6.4 & -0.017 & F1+F14-F3  \\
               & F25 & 	57.971(8) & 670.96(9) &  0.050 &  3.3 & 9.7  & 0.045 & 4.0 & 10.2 & -0.005 & \\
\hline
\end{tabular}
\end{scriptsize}
\end{center}
\tablefoot{Pulsation frequencies, amplitudes, signal-to-noise values and SigSpec significances identified from the MOST data sets from 2006 (amp$_1$, S/N$_1$,sig$_1$) and 2011/12 (amp$_2$, S/N$_2$, sig$_2$) and the CoRoT data sets from 2008 (amp$_1$, S/N$_1$,sig$_1$)and 2011/12 (amp$_2$, S/N$_2$, sig$_2$) as well as the the corresponding amplitude differences (amp$_2$ - amp$_1$). The respective last-digit errors of the frequencies computed according to Kallinger et al. (\cite{kal08}) are given in parentheses. Possible linear combinations are given in the last column (lin. combi).}
\end{table*}

\section{Spectroscopic analysis}\label{spectroscopy}
To model the frequencies detected in the light curves of HD\,261711, we 
analysed a high resolution spectrum ($R \sim$ 52\,000) obtained with the 
Robert G. Tull Coud\'e Spectrograph (TS), mounted on the 2.7-m telescope of 
Mc\,Donald Observatory. The spectrum was obtained on 2010, December 2nd, 
and covers the 3633--10849\,\AA\ wavelength range with gaps between 
the \'echelle orders at wavelengths longer than 5880\,\AA. Adopting an
exposure time of 90 minutes we obtained a signal-to-noise ratio (S/N)
per pixel, calculated over 1\,\AA\ at $\sim$5000\,\AA, of about 100.  

Bias and flat field frames were obtained at the beginning of each night, 
while several Th-Ar comparison lamp spectra were obtained each night for 
wavelength calibration purposes. The reduction was performed using the Image 
Reduction and Analysis Facility\footnote{IRAF (http://iraf.noao.edu) is 
distributed by the National Optical Astronomy Observatory, which is operated 
by the Association of Universities for Research in Astronomy (AURA) under 
cooperative agreement with the National Science Foundation.} (IRAF). The 
spectra were normalised by fitting a low order polynomial to carefully 
selected continuum points. 

We simultaneously used hydrogen lines and metallic lines to estimate the 
star's effective temperature (\Teff) and surface gravity (\logg). 
The TS spectrum, in the adopted configuration, covers fully the 
H$\gamma$ and H$\beta$ lines, but the latter cannot be used because it is 
affected by a defect of the spectrograph's imaging system. Because of gaps in the spectral orders in the red, the H$\alpha$
line is not fully covered. We normalised the H$\gamma$ line 
using the artificial flat-fielding technique described in Barklem et al. (\cite{bar02}), which has already proven to be successful with TS data (Fossati et al. \cite{fos11b}). To supplement the H$\gamma$ observation acquired at Mc\,Donald Observatory, we obtained a low resolution ($R\sim$15\,000) spectrum, covering H$\alpha$ (6470--6710\,\AA), with the 1.8-m telescope of the Dominion Astrophysical Observatory (Canada). Low resolution spectra allow a better control of the normalisation, decreasing therefore the systematics which might be introduced by considering the H$\gamma$ line, only. We reduced the DAO spectrum with IRAF in a similar way as the TS spectrum.

To compute model atmospheres of HD\,261711 we employed the \llm\ stellar 
model atmosphere code (Shulyak et al. \cite{llm}). We analysed the hydrogen lines fitting synthetic spectra, calculated with \synth\  (Kochukhov \cite{synth3}), to the observed line profiles. The hydrogen lines did not allow a very precise \Teff\ determination, leading to a best fitting temperature of 8600$\pm$400\,K; in this temperature regime hydrogen lines have little reaction to \Teff\ variations. On the other hand, we were able to better determine \logg\ to a value of 4.1$\pm$0.2.

\begin{figure}[htb]
\centering
\includegraphics[width=0.50\textwidth,clip]{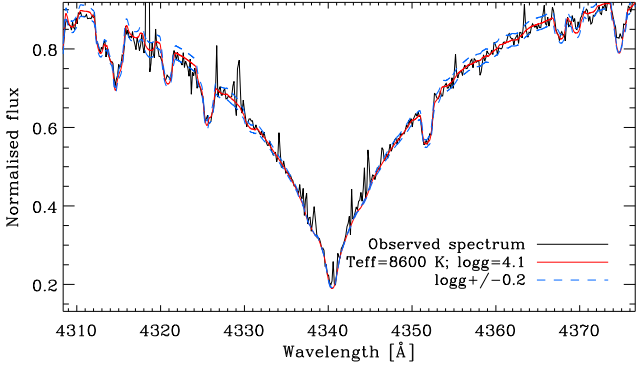}
\caption{The region of the H$\gamma$ line for HD\,261711: observed spectrum 
(black solid line), synthetic spectrum with the final adopted stellar
parameters (\Teff=8600\,K, \logg=4.1 - red thick solid line) and synthetic 
spectra calculated by increasing/decreasing \logg\ by 0.2\,dex  (blue 
dashed lines).}
\label{hgamma}
\end{figure}

\begin{figure}[htb]
\centering
\includegraphics[width=0.50\textwidth,clip]{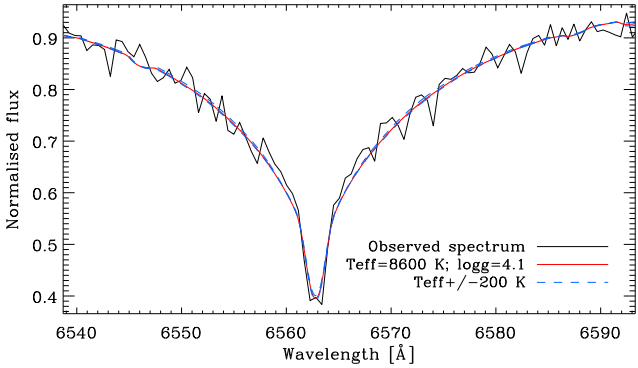}
\caption{The region of the H$\alpha$ line for HD\,261711: observed 
spectrum (black solid line), synthetic spectrum with the final adopted stellar
parameters (\Teff=8600\,K, \logg=4.1 - red thick solid line) and synthetic 
spectra calculated by increasing/decreasing \Teff\ by 200\,K (blue 
dashed lines). Note that the three synthetic spectra overlap because of the little reaction of the H$\alpha$ line to \Teff\, variations in this temperature regime.}
\label{halpha}
\end{figure}

We further constrained \Teff\ and \logg\ making use of the Fe excitation 
and ionisation equilibria. This allowed us to improve the \Teff\ determination, 
obtaining finally 8600$\pm$200\,K, and to confirm the \logg\ value of 
4.1$\pm$0.2, previously obtained with the analysis of the hydrogen lines. 
Figures~\ref{hgamma} and \ref{halpha} show a comparison between observed 
and synthetic profiles, calculated with the final adopted parameters, for 
the H$\gamma$ and H$\alpha$ lines, respectively. We also show here synthetic 
profiles calculated by increasing/decreasing \logg\ by 0.2\,dex 
(Figure~\ref{hgamma}), and \Teff\ by 200\,K (Figure~\ref{halpha}). Our fundamental parameters provide the best overall description of the available observables: hydrogen and metallic line profiles. As Table~\ref{abundance} shows, within the uncertainties we obtained the ionisation equilibrium also for elements other than Fe, such as Mg and Si.


By fitting synthetic spectra to several weakly blended lines, we measured a \vsini\ of 53$\pm$1\,\kms. The local thermodynamic equilibrium (LTE) abundance analysis was based on equivalent widths, analysed with a modified version (Tsymbal \cite{vadim}) of the {\sc WIDTH9} code (Kurucz \cite{kurucz1993a}). The relatively large \vsini\ value did not allow us to measure, with classical methods (e.g., direct integration and line profile fitting; Fossati et al. \cite{fos09}), the equivalent width of a statisically large enough number of lines. For this reason, we derived the equivalent widths of the unblended and several weakly blended lines making use of an average line profile, calculated with the least-squares deconvolution (LSD) technique (Donati et al. \cite{donati}, Kochukhov et al. \cite{kochukhov}). We iteratively fitted the observed spectrum by placing the LSD profile at the wavelength position of the various spectral lines, using as free parameter only the depth of the LSD profile. The equivalent width of the LSD profile, scaled with the fitted line depth, gives the line equivalent width. The method we adopted here to measure the equivalent widths of weakly blended lines will be described in detail in a separate work (Fossati et al., 2013, in prep.). 

We determined the microturbulent velocity (\vmic) applying the equilibrium 
between abundance and equivalent widths for all measured \ion{Fe}{i} lines, 
obtaining \vmic=3.0$\pm$0.5\,\kms. We determined the abundances of 15 
elements, listed in Table~\ref{abundance}, obtaining values comparable to 
solar and in agreement within the errors with the overall cluster 
metallicity of -0.15\,dex (Lynga \cite{lynga}). In Table~\ref{abundance} we also listed for comparison the abundances we derived for NGC\,2264\,VAS\,20 and NGC\,2264\,VAS\,87 (Zwintz et al. \cite{ZFR13}), two other members of the NGC\,2264 open cluster. The abundance pattern of the three stars is comparable within 
the uncertainties, as expected for members of the same open cluster, 
though with rather different temperatures (Fossati et al. \cite{fos11a}).

\begin{table}[ht]
\caption[ ]{LTE atmospheric abundances of HD\,216711 with the error
estimates based on the internal scatter from the number of measured lines,
$n$. }
\label{abundance}
\begin{center}
\begin{tabular}{l|cc|c|c|c}
\hline
\hline
Ion &\multicolumn{2}{|c|}{HD~216711} &  VAS\,20 &  VAS\,87 &  Sun \\      
\hline                            
    &$\log (N/N_{\rm tot})$ & $n$  &\multicolumn{3}{|c}{$\log (N/N_{\rm tot})$} \\       
\hline
\ion{C}{i }   & ~~$-$3.43$\pm$0.04 &  5 &          &   & $-$3.61~  \\                            
\ion{Na}{i}   & ~~$-$5.53$\pm$0.13 &  2 & ~~$-$5.83& $-$5.68 & $-$5.87~ \\			    
\ion{Mg}{i}   & ~~$-$4.46$\pm$0.13 &  4 & ~~$-$4.63& $-$4.55 & $-$4.44~ \\			   
\ion{Mg}{ii}  & ~~$-$4.42$\pm$0.10 &  2 &    &   & $-$4.44~ \\			   
\ion{Si}{i}   & ~~$-$4.59:         &  1 & ~~$-$4.58& $-$4.87 & $-$4.53~ \\			    
\ion{Si}{ii}  & ~~$-$4.47$\pm$0.10 &  5 & ~~$-$4.49& & $-$4.53~ \\			   
\ion{Ca}{i}   & ~~$-$5.65$\pm$0.12 & 11 & ~~$-$5.64& $-$5.71 & $-$5.70~ \\			   
\ion{Ca}{ii}  & ~~$-$5.85$\pm$0.13 &  2 &    & & $-$5.70~ \\			   
\ion{Sc}{ii}  & ~~$-$9.12$\pm$0.08 &  4 & ~~$-$8.97& $-$9.16 & $-$8.89~ \\			   
\ion{Ti}{ii}  & ~~$-$7.20$\pm$0.10 &  6 & ~~$-$7.03& $-$7.14 & $-$7.09~ \\			   
\ion{Cr}{i}   & ~~$-$6.86$\pm$0.05 &  3 & ~~$-$6.37& $-$6.39 & $-$6.40~ \\			   
\ion{Cr}{ii}  & ~~$-$6.38$\pm$0.11 &  7 & ~~$-$6.44& $-$6.06 & $-$6.40~ \\			   
\ion{Mn}{i}   & ~~$-$6.36$\pm$0.23 &  3 & ~~$-$6.36& $-$6.44 & $-$6.61~ \\			   
\ion{Fe}{i}   & ~~$-$4.62$\pm$0.20 & 45 & ~~$-$4.55& $-$4.63 & $-$4.54~ \\			   
\ion{Fe}{ii}  & ~~$-$4.61$\pm$0.14 & 27 & ~~$-$4.52& $-$4.58 & $-$4.54~ \\			   
\ion{Ni}{i}   & ~~$-$5.88$\pm$0.11 &  3 & ~~$-$5.77& $-$5.83 & $-$5.82~ \\			   
\ion{Sr}{ii}   & ~~$-$9.56:         &  1 &    &   & $-$9.17~ \\			   
\ion{Y}{ii}   & ~~$-$9.60:         &  1 & ~~$-$9.94& $-$9.66 & $-$9.83~ \\			   
\ion{Zr}{ii}  & ~~$-$9.10:         &  1 &          &   & $-$9.46~ \\			   
\ion{Ba}{ii}  & ~~$-$9.52$\pm$0.32 &  3 & ~~$-$9.29& $-$9.51 & $-$9.86~ \\			   
\hline											     %
\Teff     &\multicolumn{2}{|c|}{8600\,K}   & 6380\,K  & 6220\,K & 5777\,K  \\				
\logg     &\multicolumn{2}{|c|}{4.1~~~~~}  & 4.0  & 3.8    & 4.44~~~~\\				   
\hline											  
\end{tabular}
\end{center}
\tablefoot{For comparison purpose, columns 4,5 and 6 list respectively the abundances 
obtained by Zwintz et al. (\cite{ZFR13}) for NGC\,2264\,VAS\,20 and NGC\,2264\,VAS\,87 and for the Sun 
(Asplund et al. \cite{asp09}). }
\end{table}

\subsection{Spectral energy distribution}
\begin{figure}[htb]
\centering
\includegraphics[width=0.45\textwidth,clip]{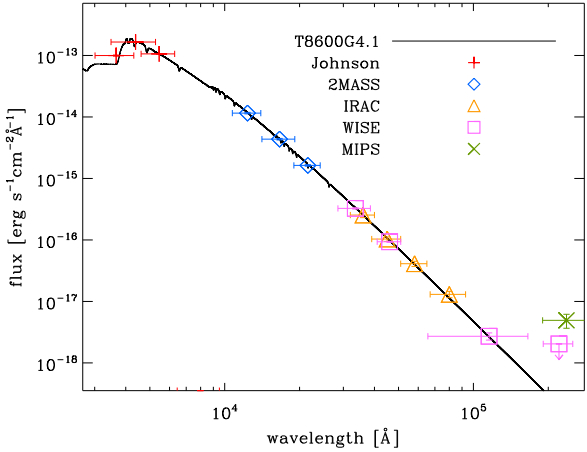}
\caption{Comparison between \llm\ theoretical fluxes (full line), calculated with the fundamental parameters derived for HD\,261711, with Johnson (crosses), 2MASS (diamonds), Spitzer IRAC (triangles), WISE (squares), and MIPS (cross) photometry converted to physical units. The horizontal lines indicate the wavelength range covered by each photometric point.}
\label{sed}
\end{figure}

Figure~\ref{sed} shows the fit of the synthetic fluxes, calculated with the fundamental parameters given before, to the observed Johnson (Mermilliod et al. \cite{mer94}), 2MASS (Zacharias et al. \cite{zac05}), Spitzer (IRAC - Sung et al. \cite{sun09}), WISE (Cutri et al. \cite{cut12}) and MIPS (Sung et al. \cite{sun09}) photometry, converted to physical units. We converted the photometry adopting the calibrations provided, respectively, by Bessel et al. (\cite{bes98}), van~der~Bliek et al. (\cite{van96}), the Spitzer IRAC instrument handbook\footnote{http://irsa.ipac.caltech.edu/data/SPITZER/docs/irac/iracinstrumenthandbook/}, Wright et al. (\cite{wri10}) and the Spitzer MIPS instrument handbook\footnote{http://irsa.ipac.caltech.edu/data/SPITZER/docs/mips/mipsinstrumenthandbook/}. Adopting the cluster distance and reddening given by Sung et al. (\cite{sun97}), we estimated a stellar radius of 1.8$\pm$0.5\,R$_\odot$. This comparison also confirms the temperature we obtained for HD\,261711.
Using the \Teff\, value of 8600\,K derived from the spectrum analysis and the radius of 1.8\Rsun, the luminosity log\,$L$/\Lsun\, for HD 261711 evaluates to 1.20 $\pm$ 0.14. 

Note that for the reddest WISE photometric point only an upper limit is available in the literature which is illustrated with small arrows in Figure~\ref{sed}. But an increased flux is also detected in the MIPS measurement which agrees to the higher flux seen in the reddest WISE photometric point. The excess in the infrared indicates that HD 261711 might still be surrounded by remnants of its birth cloud. This illustrates the relative youth of the star and, thus, supports its membership to NGC 2264.

Using the radius of 1.8 R$_\odot$ and the \vsini\ of 53 \kms, the longest possible rotation period is 1.72\,d, i.e., 0.56\,\cd (6.5\,$\mu$Hz). Hence, it is unlikely that the observed spacing between groups of frequencies corresponds to rotational splitting.
We will discuss possible explanations of the observed spacings in the following section.

\section{Asteroseismic Modeling}
\label{modeling}

HD 261711 presents an interesting pulsation spectrum in which the frequencies appear in groups. The separation in frequency between each group is equal to the large frequency spacing for a star occupying HD 261711's location in the HR diagram. This pattern is similar to that found in PMS stars HD 34282, analyzed by Casey et al. (\cite{cas13}, hereafter, CZG), and HD 144277 (Zwintz et al. \cite{zwi11b}). CZG found that the frequency separation between the weighted group-averaged frequencies corresponds to the large spacing predicted by the stellar models. 
CZG were unable to explain why the radial modes (apparently) split into groups of multiple frequencies. They considered rotational splittings, higher order $l$-values, and amplitude modulations of short-lived modes as the cause. They were able to rule out rotational splittings and higher order $l$-values but were unable to resolve the spectrum as a function of time well enough to see if any evidence for varying amplitudes existed.

Our analysis of HD 261711's oscillation spectrum follows the approach taken by CZG. We first identify the principle groups and then compute the weighted averaged frequency for each group $G_j$ according to:
\begin{equation}
G_j = \frac{\sum_{i=1}^{N_j}f_{ji}a_{ji}}{\sum_{i=1}^{N_j}a_{ji}},
\end{equation}
where $a_{ji}$ and $f_{ji}$ are the $i^{th}$ amplitude and frequency of the frequencies in the $j^{th}$ group and $N_j$ is the number of frequencies in the $j^{th}$ group. 
Values were taken from the CoRoT 2011/12 observations due to the highest accuracies in frequencies and amplitudes.
Table \ref{groups} lists the group label, the weighted average frequency, and the included frequencies (from Table \ref{freqs}).

\begin{table}[ht]
\caption[ ]{The frequency group label, the weighted average frequency, the CoRoT observed pulsation label and frequency (from Table \ref{freqs}) sorted by increasing frequency.}
\label{groups}
\begin{center}
\begin{tabular}{cccc}
\hline
\hline
\multicolumn{1}{c}{Group} &\multicolumn{1}{c}{Weighted Frequency} &  \multicolumn{1}{c}{Frequency Label}  &  \multicolumn{1}{c}{Frequency} \\                                  
    &[$\mu$Hz] &   & [$\mu$Hz] \\      
\hline	
G01 & 529.39	 & F20 & 529.39 \\
G02 & 566.92	 & F09 & 566.92 \\
G03 & 603.33	 & F11 & 592.79 \\
	 &  & 	F15	 & 602.51  \\
	 & 	 & F16	 & 606.18  \\
	 &  & 	F13	 & 614.94  \\
G04	 & 639.75	 & F12 & 633.58 \\
	 &  & 	F19	 & 634.35  \\
	 &  & 	F07	 & 637.28  \\
	 &  & 	F10	 & 640.81  \\
	 &  & 	F06	 & 641.86  \\
	 &  & 	F17	 & 646.89  \\
G05	 & 671.94	 & F25 & 670.96 \\
	 &  & 	F02	 & 671.61  \\
	 &  & 	F05	 & 671.94  \\
	 &  & 	F18	 & 673.97  \\
	 &  & 	F08	 & 676.26  \\
	 &  & 	F04	 & 678.03  \\
G06	 & 711.76	 & F14 & 709.27 \\
	 & &  	F24 & 	709.91  \\
	 &  & 	F03	 & 711.11  \\
	 &  & 	F21 & 	711.38  \\
	 &  & 	F01	 & 711.78  \\
G07	 & 749.59 & 	F22 & 749.59 \\
G08 & 	970.34 & 	F23 & 970.34 \\
\hline									  
\end{tabular}
\end{center}
\end{table}

\begin{figure}[htb]
\centering
\includegraphics[width=0.5\textwidth,clip]{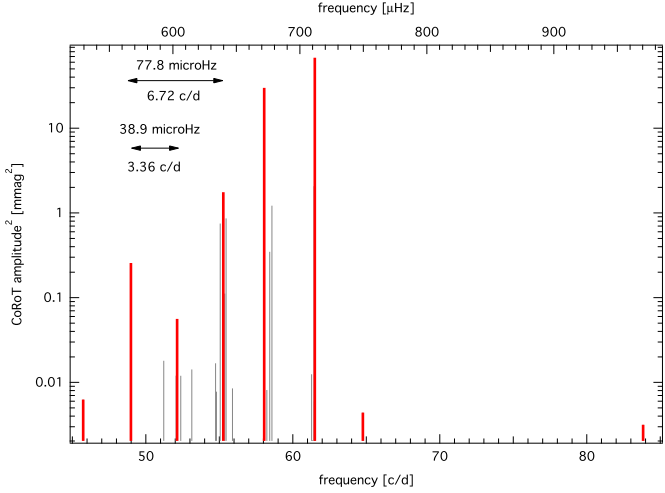}
\caption{Weighted (red thick lines) and observed (grey thin lines) pulsation frequencies with squared amplitudes of HD 261711 where the characteristic mean observed spacing of about 3.36\,\cd\, (38.9\,$\mu$Hz) and twice its value, i.e., 6.72 \,\cd\, (77.8\,$\mu$Hz), are indicated.}
\label{fspec}
\end{figure}

Figure \ref{fspec} shows the observed 25 pulsation frequencies as thin grey lines and the weighted average frequencies as thick red lines with squared amplitudes. The observed average characteristic spacing is around 3.36\,\cd\, (38.9\,$\mu$Hz) which is indicated by arrows together with twice the observed spacing of about 6.72 \,\cd\, (77.8\,$\mu$Hz).

\begin{figure}[htb]
\centering
\includegraphics[width=0.45\textwidth,clip]{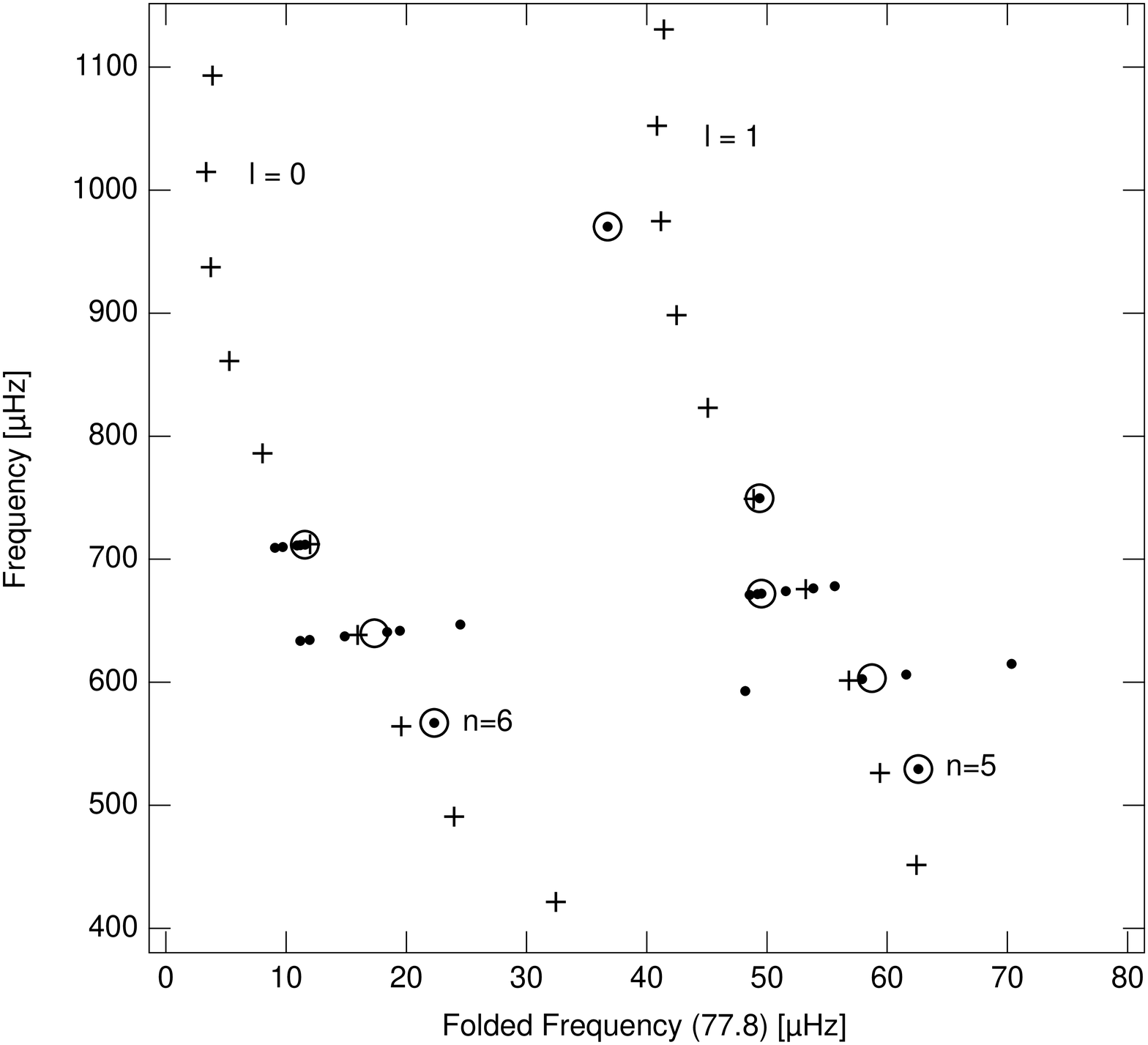}
\caption{Echelle diagram showing the CoRoT 2011/12 observations (small filled circles), the weighted average frequencies (large open circles), and the selected best model fit (plus symbols) with an identification of the $n$ values from the models.}
\label{echelle}
\end{figure}

In Figure \ref{echelle} we plot the observed frequencies and the weighted-averaged frequencies in an echelle diagram (frequency versus frequency modulo a folding frequency). The folding frequency is 77.8\,$\mu$Hz and was chosen to produce vertically aligned ridges of modes, a signature of $p$-modes. The $l = 0$ and 1 $p$-mode frequencies of our best model fit to the weighted averaged frequencies, to be discussed below, are also shown in Figure \ref{echelle}. 

After failing to find a model whose frequencies fit all the observed frequencies, we attempted to fit a model to just the eight weighted averaged frequencies. We used the usual $\chi^{2}$ fitting procedure, first described in Guenther and Brown (\cite{gue04}). The match between the observed and model spectra is quantified by the following $\chi^{2}$ equation: 
\begin{equation}
\chi^{2} = \frac{1}{N}{\sum_{i=1}^{N}} \frac{(f_{obs,i} - f_{mod,i})^2}{\sigma{^2}_{obs,i} + \sigma{^2}_{mod,i}}
\end{equation}
where $N$ is the number of weighted averaged frequencies, $f_{obs,i}$ and $f_{mod,i}$ are the $i^{th}$ weighted averaged and model frequencies, respectively, and $\sigma{^2}_{obs,i}$ and $\sigma{^2}_{mod,i}$ are the $i^{th}$ weighted average and model frequency uncertainties. Here we assume the model frequency uncertainties are small compared to the weighted averaged frequency uncertainties. For simplicity of modeling and because we do not have a formal explanation for the cause of the splittings within each group, we set the uncertainty of all the weighted averaged frequencies to $\pm 1\mu$Hz. 

The model frequencies were taken from the same PMS model grid as used in Guenther et al. (\cite{gue09}). The model grid was constructed using the YREC stellar evolution code (Demarque et al. \cite{dem08}). Solar metallicity (Z = 0.02) models between 1.00 and 5.00 \Msun, in increments of 0.01 \Msun, were used. The PMS models, themselves, lie along evolutionary tracks that start on the Hayashi track and end on the ZAMS where nuclear burning begins. The non-adiabatic stellar pulsation program by Guenther (\cite{gue94}) was used to calculate the adiabatic $p$-mode frequency spectrum of each model. Radial orders $n$ = 0 to 30 and azimuthal orders $l$ = 0 to 3 were compared to the weighted averaged frequencies and the models with the lowest $\chi^{2}$ were identified.

\begin{figure}[htb]
\centering
\includegraphics[width=0.45\textwidth,clip]{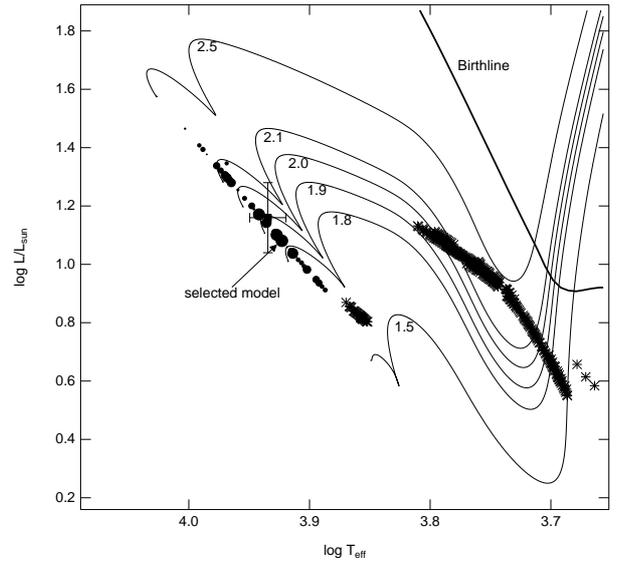}
\caption{HR-diagram showing the lowest $\chi^{2}$ models (crosses). The models with low $\chi^{2}$ and lying near the star's location in the HR-diagram are shown with filled circles whose radii are inversely proportional to $\chi^{2}$. Also shown are PMS evolutionary tracks with the indicated masses (\Msun) and the birthline. HD 261711's observational HR-diagram position including uncertainties was derived using the spectroscopically determined value of log\Teff\, and the asteroseismic radius of 1.65\Rsun.}
\label{HRD}
\end{figure}

Figure \ref{HRD} shows the regions in the HR-diagram where the models have the lowest $\chi^{2}$ ($<$ 10). The models that lie close to HD 261711's position in the HR-diagram derived from our observations are represented by circle sizes that are inversely proportional to $\chi^{2}$ with the largest circles corresponding to models whose oscillation spectra best match the weighted averaged frequencies. Two regions of models were found to fit the weighted averaged frequencies of the observed oscillation spectrum, one near the ZAMS and the other in a region below the birthline. The models near the birthline correspond to fits in which the weighted averaged frequencies correspond to purely radial or purely $l$ = 1 $p$-modes. The best fitting models near the ZAMS correspond to fits in which the weighted averaged frequencies correspond to both $l$ = 0 and 1 $p$-modes. At this time we do not consider further the low $\chi^{2}$ models near the birthline because they do not match the starÕs location in the HR-diagram derived from observations. 

The models that do match the weighted averaged frequencies and lie within the observed HR-diagram position of HD 261711 have masses from 1.8 \Msun to 1.9 \Msun and ages from 10 to 11 Myrs (from birthline).

The asteroseismic observations and the spectroscopic observations are consistent; hence, it appears that the weighted averaged frequencies are related to the $l$ = 0 and 1 $p$-modes. 
If our $l$-value identifications are correct then, it is not possible for the radial modes to be split by rotation; hence, rotation as a cause of the splitting is ruled out. Although we cannot completely rule out the possibility that the modes are $l$ = 1 and 2, or 2 and 3 but we note that these fits have higher $\chi^{2}$ than those shown here. 

\begin{figure}[htb]
\centering
\includegraphics[width=0.45\textwidth,clip]{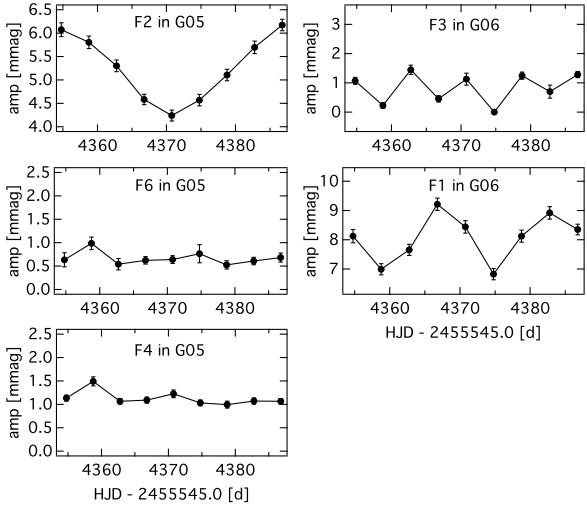}
\caption{Modulation of the amplitudes of frequencies F2 (top left), F6 (middle left) and F4 (bottom left) of group G05 and F3 (top right) and F1 (middle right) of group G06 when splitting the CoRoT 2011/12 data sets in nine subsets of $\sim$5\,d length. Amplitude errors are computed according to Kallinger et al. (\cite{kal08}).}
\label{ampvar}
\end{figure}

To see if the mode amplitudes varied during the period of observation, which could introduce spurious peaks about the parent frequency during the Fourier extraction of the star's oscillation spectrum, we split the data up into smaller subsets.  Using the longest data set, i.e. the CoRoT 2011/12 observations, and splitting them in half, i.e., first $\sim$20\,d and second $\sim$20\,d with one day overlap, showed no significant amplitude changes in the modes. But when we split the  same data set into nine subsets of five days each (to maintain some frequency resolution) with one day overlaps, we were able to resolve amplitude variations among some of the modes in G05 (three modes) and G06 (two modes), see Figure \ref{ampvar}. The amplitude modulations could be caused by nonlinear coupling of the modes to each other or some other periodic effect within the star or they could simply be an instrumental effect. Regardless, the observed amplitude modulation does offer a simple explanation for the apparent multiple peaks in the Fourier spectrum. The simplest interpretation of the oscillation spectrum remains, i.e., that they are radial and first order $p$-modes.

The oscillation spectrum of one of the best fitting models (shown in Figure \ref{HRD}) is plotted in Figure \ref{echelle} along with the observations. The selected best fitting model has an age of 11 Myr and a radius of 1.65\,\Rsun\, in agreement with the radius derived from spectral energy distribution (SED) fitting. 
The echelle diagrams of other best fitting models are similar at the level of our assumed uncertainties. The model fit shows that the weighted averaged frequencies for HD 261711 align along two vertical ridges in the echelle diagram corresponding to $l$ = 0 and 1 $p$-modes of $n$ = 6 and 5, respectively. The folding frequency of the echelle diagram is approximately equal to the large frequency spacing. Our result is consistent with Guenther et al. (\cite{gue09}) who first studied this star using just the two frequencies observed by MOST (see Table \ref{freqs}).

Comparing the best model fit to the observed frequencies and to the weighted averaged frequencies, it appears that the model fit frequencies are better aligned to the center of the groups of frequencies than the weighted averaged frequencies. 
Unlike the previously studied HD 34282 (CZG), HD 261711 shows both radial and nonradial $p$-modes. Also, unlike HD 34282, all the observed frequencies for HD 261711 fall well below the acoustic cutoff frequency. The observed frequencies for HD 34282 appear to extend right up to the acoustic cutoff frequency (CZG).

\section{Evolutionary stage of HD 261711}

Close to the ZAMS, the atmospheric properties (\Teff, \logg, luminosity and mass) of pre- and (post-) main sequence stars are quite similar (Marconi \& Palla \cite{mar98}). A distinction between evolutionary stages just from the stars' positions in the HR-diagram is not possible because pre- and post-main sequence evolutionary tracks intersect close to the ZAMS (Breger \& Pamyatnykh \cite{bre98}). 
The main difference between PMS stars and their more evolved counterparts lies in their interior structures which can be investigated using asteroseismology.

\begin{figure}[htb]
\centering
\includegraphics[width=0.49\textwidth,clip]{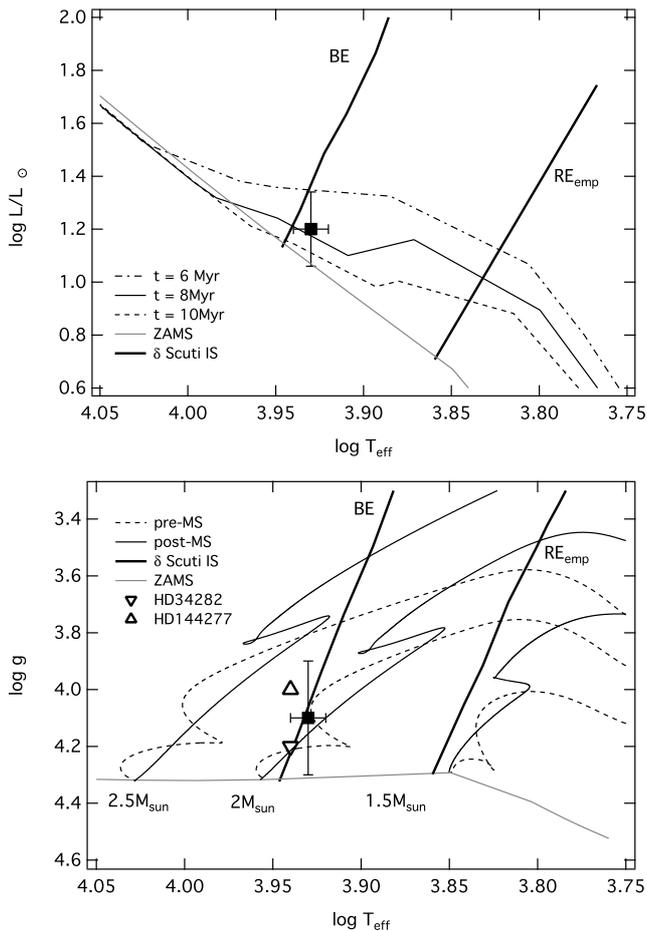}
\caption{Position of HD 261711 in the HR-diagram in comparison with isochrones (top panel) and evolutionary tracks (bottom panel). In both panels the ZAMS is indicated by a grey, solid line and the borders of the \dsct\ instability strip are marked with thick solid lines. The position of HD 261711 is given with a filled square. Top panel: Isochrones with ages of 6, 8 and 10\,Myr are indicated with a dash-dotted, solid and dashed line. Bottom panel: Pre-main sequence evolutionary tracks for 1.5, 2.0 and 2.5\Msun\ are shown with dashed lines, (post-) main sequence evolutionary tracks for the same masses with solid lines.  The open symbols mark the positions of HD 34282 (down triangle) and HD 144277 (up triangle). }
\label{obshrd}
\end{figure}

The bottom panel of Figure \ref{obshrd} shows the HR-diagram position of HD 261711 together with the pre- and post-main sequence evolutionary tracks for 1.5, 2.0 and 2.5 \Msun, and the borders of the \dsct\, instability strip. Pre- and post-main sequence evolutionary tracks (dashed and solid lines) were taken from D. Guenther (private comm.) using the YREC evolution code (Demarque et al. \cite{dem08}) with physics described in Guenther et al. (\cite{gue09}). The borders of the classical \dsct\, instability strip (i.e., the general blue edge, BE, and the empirical red edge, RE$_{\rm emp}$, thick solid lines) are from Breger \& Pamyatnykh (\cite{bre98}). The ZAMS is indicated by a grey line. The PMS \dsct\, instability strip coincides well with the classical \dsct\, instability strip (Zwintz \cite{zwi08}), hence it is sufficient to plot one of the two.
HD 261711 is located at the hot border of the instability region and close to the intersecting evolutionary tracks of 2.0 \Msun. For comparison the positions of the two similar objects, HD 34282 (Casey et al. \cite{cas13}) and HD 144277 (Zwintz et al. \cite{zwi11b}) are given. From this plot alone, no clear distinction of the evolutionary stage for HD 261711 (and also for HD 34282 and HD 144277) can be drawn.

Independent information about the age of the star can resolve the evolutionary state. In the case of HD 261711, we can make use of two additional pieces of information, the cluster membership and the age derived from asteroseismology.

HD 261711 is a very likely member of the young open cluster NGC 2264 according to its proper motion values of {\mbox -3.00 $\pm$ 2.60 mas/yr} in right ascension and -4.20 $\pm$ 2.40 mas/yr in declination (Hog et al. \cite{hog00}). The corresponding values for NGC 2264 are -2.70 $\pm$ 0.25 mas/yr and -3.50 $\pm$ 0.26 mas/yr (Kharchenko et al. \cite{kha01}). We determined a radial velocity of 14 $\pm$ 2 \kms\, for HD 261711 from our spectrum which agrees also well to the cluster radial velocity of 17.68 $\pm$ 2.2 \kms\, (Kharchenko et al. \cite{kha05}).
NGC 2264 has a maximum age of 10 million years (e.g., Sung et al. \cite{sun04}, Sagar et al. \cite{sag86}). Hence, all stars of spectral types later than A0 that lie above the ZAMS must be in the PMS evolutionary phase. Therefore, HD 261711 being a cluster member cannot be an evolved \dsct\, star.

The upper panel of Figure \ref{obshrd} shows the HR-diagram position of HD 261711 together with isochrones of 6 (dash-dotted line), 8 (solid line) and 10 (dashed line) million years (Demarque et al. \cite{dem08}). For the isochrones, we adopted the slightly less than solar metallicity reported for NGC 2264, i.e., [Fe/H] = -0.15. Under the assumption that HD 261711 is a member of NGC 2264, it can be derived from Figure \ref{obshrd} that its age is about 8 $\pm$ 2 million years.

From our asteroseismic analysis, we also find that the best fitting models put the star relatively close to the ZAMS (see Section \ref{modeling} and Figure \ref{HRD}) and yield an age of 10 million years. This again is an argument that HD 261711 cannot be an evolved \dsct\, star, and the asteroseismic age fits well to the age derived from the isochrones using the star's membership in NGC 2264.

Many PMS stars show emission lines in their spectra or an increased flux in the infrared originating from the remnants of their birth clouds. In the case of NGC 2264, the dense cloud of material apparently lies behind the stars. This is supported by the presence of only small amounts of reddening across the cluster reported in the literature (e.g., Sung et al. \cite{sun97}). The SED of HD 2617211 (Figure \ref{sed}) shows a slightly increased flux in the infrared that might be attributed to the presence of circumstellar material along the line of sight.
We did not detect emission lines in our spectrum. This is not too surprising because several other members of NGC 2264 with properties similar to HD 261711 have been reported not to show emission in their spectra, e.g., NGC 2264 VAS 20 and NGC 2264 VAS 87 (Zwintz et al. \cite{ZFR13}) or HD 261230 and HD 261387 (Zwintz et al. \cite{zwi09}). 

Putting all this information together we conclude that HD 261711 is either still in its PMS phase just before arrival on the ZAMS or it is already a young main sequence star which has just started to burn hydrogen in its core.

\section{Summary}

High-precision time series photometry from the MOST and CoRoT satellites was obtained for HD~261711 in 2006, 2008 and 2011/12. Using these four data sets, 25 \dsct-type pulsation frequencies have been discovered that lie between 45 and 84\,\cd, i.e., between about 520 and 972\,$\mu$Hz.

The observed frequencies are found in groups with a characteristic spacing of about 3.36\,\cd\ (38.9\,$\mu$Hz). This pattern resembles two other cases reported earlier, i.e., HD 34282 (CZG \cite{cas13}) and HD~144277 (Zwintz et al. \cite{zwi11b}). 

Using dedicated high-resolution spectroscopy we derived the atmospheric parameters and chemical abundances for HD~261711. The star has an effective temperature of 8600\,$\pm$ 200\,K which corresponds well to the reported spectral type of A2 found in the literature and to one of the best asteroseismic model fitting regions presented here.
\logg\ is 4.1 $\pm$ 0.2 and \vsini\ was found to be 53\,$\pm$\,1\,\kms. Chemical abundances were derived for 15 elements, are comparable to the solar values and agree within the errors to the overall cluster metallicity.  

As it was not possible to fit asteroseismic models to all observed frequencies, we fitted models to the weighted averaged frequencies of the eight observed groups. Two regions of models in the HR-diagram were found to fit these eight frequencies best: one is near the birthline and includes either purely radial or purely $l$ = 1 $p$-modes; the other is close to the ZAMS and the frequencies correspond to $l$ = 0 and $l$ = 1 $p$-modes with $n$ values of 6 and 5, respectively. The latter model also matches the location of HD 261711 in the HR-diagram obtained from spectroscopy. Therefore, we interpret HD 261711 to show $l$ = 0 and $l$ = 1 $p$-modes, to have a mass between 1.8 and 1.9\,\Msun\, and an age between 10 and 11 million years. 

We can confirm the membership of HD 261711 to the young cluster NGC 2264 as (i) its chemical abundances agree with the overall cluster metallicity, (ii) the \Teff\, and \logg\ values derived from spectroscopy yield HD 261711 to be a hot A star close to the ZAMS, (iii) its radial velocity agrees to the cluster values, (iv) there is an indication for an infrared excess in the SED, and (v) the age determined from asteroseismology coincides well within the uncertainties to the overall cluster age.

We now have three stars Ð HD 34282 (CZG \cite{cas13}), HD 144277 (Zwintz et al. \cite{zwi11b}) and HD 261711 Ð that show a similar $p$-mode spectra pattern and fit our models. The appearance of multiple frequencies in groups could be a consequence of the frequency extraction process which does not allow for varying amplitude modes. We do observe amplitude modulation in some of the modes. At this time, though, we cannot say whether the amplitude modulation is intrinsic, presumably caused by some nonlinear coupling, or an instrumental effect.

\begin{acknowledgements}
KZ receives a Pegasus Marie Curie Fellowship of the Research Foundation - Flanders (FWO). This investigation has been supported by the Austrian Fonds zur F\"{o}rderung der wissenschaftlichen Forschung through project P 21830-N16 (PI: M. Breger). 
TR acknowledges partial financial support from Basic Research Program of the Russian Academy of Sciences ``Non-stationary phenomena in the Universe''.
DBG, JM, AFJM and SMR acknowledge the funding support of the Natural Sciences and Engineering Research Council (NSERC) of Canada.
RK and WWW are supported by the Austrian Fonds zur F\"{o}rderung der wissenschaftlichen Forschung (P22691-N16) and by the Austrian Research Promotion Agency-ALR.
Spectroscopic data were obtained with the 2.7-m telescope at Mc Donald Observatory, Texas, US and at the Dominion Astrophysical Observatory, Herzberg Institute of Astrophysics, National Research Council of Canada.
\end{acknowledgements}


\begin{thebibliography}{}
\bibitem[2009]{asp09} Asplund, M., Grevesse, N., Sauval, A. J., Scott, P. 2009, ARA\&A, 47, 481
\bibitem[2009]{auv09} Auvergne, M., Baudin, P., Boisnard, L., et al. 2009, \aap, 506, 411
\bibitem[2006]{bag06} Baglin, A. 2006, The CoRoT mission, pre-launch status, stellar seismology and planet finding (M. Fridlund, A. Baglin, J. Lochard and L. Conroy eds, ESA SP-1306, ESA Publication Division, Noordwijk, The Netherlands)
\bibitem[2002]{bar02} Barklem, P. S., Stempels, H. C., Allende Prieto, C., et al. 2002, \aap, 385, 951
\bibitem[1998]{bes98} Bessel, M.S., Castelli, F. \& Plez, B. 1998, \aap, 333, 231
\bibitem[1972]{bre72} Breger, M. 1972, \apj, 171, 539
\bibitem[1993]{bre93} Breger, M. 1993, \aap, 271, 482
\bibitem[1998]{bre98} Breger, M. \& Pamyatnykh, A. A. 1998, \aap, 332, 958
\bibitem[2008]{bre08} Breger, M., Lenz, P. \& Pamyatnykh, A. A. 2008, CoAst, 157, 56
\bibitem[2009]{bre09} Breger, M., Lenz, P. \& Pamyatnykh, A. A. 2009, \mnras, 396, 291
\bibitem[2011]{bre11} Breger, M., Balona, L., Lenz, P., et al. 2011, \mnras, 414, 1721
\bibitem[2011]{bou11} Bouabid, M.-P., Montalb\'an, J., Miglio, A., et al. 2001, \aap, 531, 145
\bibitem[1933a]{carroll33a} Carroll, J. A. 1933a, \mnras, 93, 478
\bibitem[1933b]{carroll33b} Carroll, J. A. 1933b, \mnras, 93, 680
\bibitem[2012]{cas13} Casey, M., Zwintz, K., Guenther, D. B., et al. 2013, \mnras, 428, 2596
\bibitem[2012]{cut12} Cutri, R. M., Wright, E. L., Conrow, T., et al. 2012, WISE All-Sky Data Release, VizieR On-line Data Catalog: II/311, http://cdsads.u-strasbg.fr/abs/2012yCat.2311....0C
\bibitem[2008]{dem08} Demarque, P., Guenther, D. B., Li, L. H., Mazumdar, A., \& Straka, C. W. 2008, \apss, 316, 31
\bibitem[1997]{donati} Donati, J.-F., Semel, M., Carter, B.~D., Rees, D.~E., \& Collier Cameron, A. 1997, \mnras, 291, 658
\bibitem[1990]{dravins90} Dravins, D., Lindegren, L. \& Torkelsson, U. 1990, \aap, 237, 137
\bibitem[2007]{fos07} Fossati, L., Bagnulo, S., Monier, R., et al. 2007, \aap, 476, 911 
\bibitem[2009]{fos09} Fossati, L., Ryabchikova, T., Bagnulo, S., et al. 2009, \aap, 503, 945
\bibitem[2011a]{fos11a} Fossati, L, Folsom, C. P., Bagnulo, S., et al. 2011a, \mnras, 413, 1132
\bibitem[2011b]{fos11b} Fossati, L., Ryabchikova, T., Shulyak, D. V., et al. 2011b, \mnras, 417, 495
\bibitem[2008]{glazunova08} Glazunova, L.~V., Yushchenko, A.~V., Tsymbal, V.~V., et al. 2008, \aj, 136, 1736 
\bibitem[1994]{gue94} Guenther, D. B. 1994, \apj, 422, 400
\bibitem[2004]{gue04} Guenther, D. B., \& Brown, K. I. T. 2004, \apj, 600, 419
\bibitem[2009]{gue09} Guenther, D. B., Kallinger, T., Zwintz, K., et al. 2009, \apj, 704, 1710
\bibitem[2008]{har08} Hareter, M., Reegen, P., Kuschnig, R., et al. 2008, CoAst, 156, 48
\bibitem[2000]{hog00} Hog, E., Fabricius, C., Makarov, V. V., et al. 2000, \aap, 355, 27
\bibitem[2008]{kal08} Kallinger, T., Reegen, P. \& Weiss, W. W. 2008, \aap, 481, 571
\bibitem[2001]{kha01} Kharchenko, N. V. 2001, Kinematics and Physics of Celestial Bodies, 17, 409
\bibitem[2005]{kha05} Kharchenko, N.N., Piskunov, A. E., Roeser, S., Schilback, E., Scholz, R. D. 2005, \aap, 438, 1163
\bibitem[2007]{synth3} Kochukhov, O. 2007, Spectrum synthesis for magnetic, chemically stratified stellar atmospheres, in Physics of Magnetic Stars, eds. I.~I.~Romanyuk, D.~O.~Kudryavtsev, O.~M.~Neizvestnaya, \& V.~M.~Shapoval (Special Astrophysical Observatory, RAS), 109 
\bibitem[2010]{kochukhov} Kochukhov, O., Makaganiuk, V. \& Piskunov, N. 2010, \aap, 524, A5
\bibitem[1999]{vald2} Kupka, F., Piskunov, N., Ryabchikova, T. A., Stempels, H. C., Weiss, W. W. 1999, \aaps, 138, 119
\bibitem[1993]{kurucz1993a} Kurucz, R. 1993, ATLAS9: Stellar Atmosphere Programs and 2 km/s grid.~Kurucz CD-ROM No.~13 (Cambridge: Smithsonian Astrophysical Observatory)
\bibitem[1997]{kus97} Kuschnig, R., Weiss, W. W., Gruber, R., Bely, P. Y., Jenkner, H. 1997, \aap, 328, 544
\bibitem[2005]{len05} Lenz, P. \& Breger, M. 2005, CoAst, 146, 53
\bibitem[1987]{lynga} Lynga, G. 1987, Catalogue of Open Cluster Data, 5th Ed., Lund Observatory, Sweden
\bibitem[1998]{mar98} Marconi, M. \& Palla, F., 1998, \aj, 507, L141
\bibitem[2011]{mash11} Mashonkina, L., Gehren, T., Shi, J.-R., Korn, A. J., Grupp, F. 2011, \aap, 528, 87
\bibitem[1994]{mer94} Mermilliod, J.-C. \& Mermilliod, M. 1994, Catalogue of Mean UBV Data on Stars, VI, 1387 pp. Springer-Verlag Berlin Heidelberg New York
\bibitem[1985]{moon} Moon, T. T. \& Dworetsky, M. M. 1985, \mnras, 217, 305 
\bibitem[1995]{vald1} Piskunov, N. E., Kupka, F., Ryabchikova, T. A., Weiss, W. W., \& Jeffery, C. S. 1995, \aaps,  112, 525
\bibitem[2006]{ree06} Reegen, P., Kallinger, T., Frast, D. et al. 2006, \mnras, 367, 1417
\bibitem[2007]{ree07} Reegen, P. 2007, \aap, 467, 135
\bibitem[2002a]{reiners1} Reiners, A. \& Schmitt, J. H. M. M. 2002a, \aap, 384, 155
\bibitem[2002b]{reiners2} Reiners, A. \& Schmitt, J. H. M. M. 2002b, \aap, 393, L77
\bibitem[2006]{rip06} Ripepi, V., Bernabei, S., Marconi, M., et al. 2006, \aap, 449, 335
\bibitem[2011]{rip11} Ripepi, V., Cusano, F., di Criscienzo, M., et al. 2011, \mnras, 416, 1535
\bibitem[1999]{vald3} Ryabchikova, T.~A., Piskunov, N.~E., Stempels, H.~C., Kupka, F., \& Weiss, W.~W. 1999, Phis. Scr., T83, 162
\bibitem[1986]{sag86} Sagar, R., Piskunov, A. E., Miakutin, V. I., Joshi, U. C. 1986, \mnras, 220, 383
\bibitem[2004]{llm} Shulyak, D., Tsymbal, V., Ryabchikova, T., St\"utz\, Ch., Weiss, W. W. 2004, \aap, 428, 993
\bibitem[2005]{ski05} Skiff, B. A. 2005, General Catalog of Stellar Spectral Classifications, Lowell Observatory
\bibitem[1997]{sun97} Sung, H., Bessel, M. S. \& Lee, S.-W. 1997, \aj, 114, 2644
\bibitem[2004]{sun04} Sung, H., Bessel, M. S., \& Chun, M.Y. 2004, \aj, 128, 1684
\bibitem[2009]{sun09} Sung, H., Sauffer, J. R., Bessel, M. S. 2009, \aj, 138, 1116
\bibitem[1996]{vadim} Tsymbal, V. V. 1996, in ASP Conf.~Ser.~108, Model Atmospheres and Spectral Synthesis, ed. S.~J., Adelman,	F., Kupka, \& W.~W., Weiss, 198
\bibitem[1996]{van96} van~der~Bliek, N.~S., Manfroid, J. \& Bouchet, P. 1996, \aaps, 119, 547
\bibitem[1993]{vanhamme93} Van Hamme, W. 1993, \aj, 106, 20 
\bibitem[2003]{wal03} Walker, G., Matthews, J. M., Kuschnig, R., et al. 2003, \pasp, 115, 1023
\bibitem[2006]{wei06} Weiss, W. 2006, in Proc. The CoRoT Mission Pre Launch Status, Stellar Seismology and Planet Finding, ed. M. Fridlund, A. Baglin, J. Lochard, \& L. Conroy (ESASP 1306; Noordwijk, The Netherlands: ESA Publications Division), 93
\bibitem[2002]{wei02} Weisskopf, M. C., Brinkman, B., Canizares, S., et al. 2002, \pasp, 114, 1
\bibitem[2004]{wer04} Werner, M. W., Roellig, T. L., Low, F. J., et al. 2004, \apjs, 154, 1
\bibitem[2010]{wri10} Wright, E. L., Eisenhardt, P. R. M., Mainzer, A. K., et al. 2010, \aj, 140, 1868
\bibitem[2005]{zac05} Zacharias, N., Monet, D.~G., Levine, S.~E., Urban, S.~E., Gaume, R. \& Wycoff, G.~L 2005, BAAS, 36, 1418
\bibitem[2008]{zwi08} Zwintz, K. 2008, \apj, 673, 1088
\bibitem[2009]{zwi09} Zwintz, K., Hareter, M., Kuschnig, R., et al. 2009, \aap, 502, 239
\bibitem[2011a]{zwi11a} Zwintz, K., Kallinger T., Guenther, D. B., et al. 2011a, \apj, 729, 20
\bibitem[2011b]{zwi11b} Zwintz, K., Lenz, P., Breger, M., et al. 2011b, \aap, 533, 133
\bibitem[2013]{ZFR13} Zwintz, K., Fossati, L., Ryabchikova, T. A., et al. 2012, \aap, in print, (2013arXiv1301.0991Z)
\end{thebibliography}
\end{document}